\documentclass[twocolumn,showpacs]{revtex4}

\usepackage{graphicx}
\usepackage{dcolumn}
\usepackage{amsmath}

\begin{document}

\title[Short Title]{Conductance via Multi orbital Kondo Effect in Single Quantum Dot}

\author{Rui Sakano}
\author{Norio Kawakami}
\affiliation{
Department of Applied Physics,
Osaka University,  Suita, Osaka 565-0871, Japan\\
}
\date{\today}

\begin{abstract}
We study the Kondo effect in a single quantum dot system with two or three orbitals by using the Bethe-ansatz exact solution at zero temperature and the non-crossing approximation at finite temperatures. 
For the two-orbital Kondo effect, the conductance is shown to be constant at absolute zero in any magnetic fields, but decrease monotonically with increasing fields at finite temperatures. In the case with more orbitals, the conductance increases at absolute zero, while it features a maximum structure as a function of the magnetic field at finite temperatures. We discuss how these characteristic transport properties come from  the multi orbital Kondo effect in magnetic fields.
\end{abstract}

\maketitle

\section{Introduction}
The Kondo effect is a typical electron correlation problem, which was originally discovered in dilute magnetic alloys as a resistance minimum phenomenon \cite{Kondo,book:hewson}.  Although the essence of the Kondo effect was already clarified, it is quite interesting to systematically study the Kondo effect in artificial systems. Recently, the Kondo effect observed in quantum dot systems,
 which have rich tunable parameters \cite{Kouwenhoven,Reimann}, has attracted considerable attention \cite{DGG,pap:Cronenwett,pap:Schmid}. This has stimulated intensive investigations on electron correlations. Indeed, a lot of the work have been done both theoretically and experimentally.

Among intensive studies, the Kondo effect due to the orbital degrees of freedom is of particular interest, where effective orbitals are given by a highly symmetric shape of the dot, multiply coupled  quantum dots, etc \cite{pap:Tarucha,pap:Moriyama,pap:Jarillo-Herrero1,pap:Bellucci,pap:Cobden,pap:Asano,pap:Park}.
In such systems, not only the ordinary spin Kondo effect \cite{pap:TK,pap:Glazman} but also the orbital Kondo effect due to the interplay between orbital and spin degrees of freedom occurs; the Kondo effect in multiple-dot systems \cite{pap:Wilhelm,pap:Borda,pap:Li,pap:Holleitner,pap:Galpin,pap:Sakano,pap:LopezSU4DD,pap:Chud,pap:Kuzmenkomd,pap:Lipinski}, the
singlet-triplet Kondo effect \cite{pap:Sasaki,pap:KKikoin,pap:Pustilnik,pap:EtoS-D}, and the SU(4) Kondo effect in single-dot systems \cite{pap:Sasaki2,pap:Zhuravlev,pap:EtoSU4,pap:Herrero,pap:Choi}.

In this paper, we focus on the Kondo effect in a single-dot system with multiple orbitals, and explore characteristic transport properties at zero and finite temperatures. To this end, we use the Bethe-ansatz exact solution of the impurity Anderson model and the non-crossing approximation (NCA). To be specific, we deal with a quantum dot system with two or three orbitals as a typical example of the multi-orbital Kondo effect.

Experimentally, the two-orbital Kondo effect
(SU(4) Kondo effect)  in single-dot systems has already been observed in transport properties, where the conductance shows similar field-dependence to the usual SU(2) spin Kondo effect: in both cases the conductance is monotonically suppressed in the presence of magnetic fields\cite{pap:Sasaki2,pap:Herrero}. One of the main purposes of the preset paper is to demonstrate that 
although  the field-dependent conductance shows analogous behavior, 
its origin is different from each other: 
 the decrease of the conductance in the two-orbital case is
due to the decrease of the effective Kondo temperature, 
in contrast to the ordinary SU(2) Kondo effect. Moreover, 
we show that in three- or more-orbital cases, the
conductance even features a maximum structure in magnetic fields, which
is not expected in the ordinary spin Kondo effect.

This paper is organized as follows. In Sec. \ref{sc:model}, we introduce a generalized impurity Anderson model for our multi-orbital quantum dot  system, and outline how to treat transport properties by means of the Bethe-ansatz exact solution at absolute zero and the NCA method at finite temperatures. 
We investigate the two-orbital Kondo effect in Sec. \ref{sec:two-orbital},  and then move to the case with more orbitals in Sec. \ref{sec:three-orbital} by taking the three-orbital Kondo effect as an example. We discuss characteristic magnetic-field dependence of the conductance due to orbital effects at zero and finite temperatures. A brief summary is given in Sec. \ref{sec:conc}, where the multi-orbital Kondo effects are compared with the ordinary single-orbital spin Kondo effect.

\section{model and method} \label{sc:model}

\subsection{Multi orbital quantum dot}

\begin{figure}[bth]
\includegraphics[width=4.5cm]{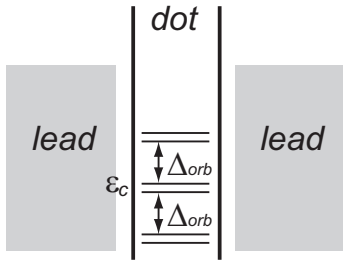}
\caption{Energy-level scheme of a single quantum dot with three orbitals coupled to two leads.}
\label{fig:schematic}
\end{figure}

Let us consider a single quantum dot system with $N$-degenerate orbitals in 
equilibrium state.
As shown in Fig. \ref{fig:schematic}, 
we assume that each energy-level splitting between the orbitals 
is $\Delta_{orb}$ in the 
presence of magnetic field and the Zeeman splitting 
is much smaller than it, so that we can ignore the Zeeman effect: 
the degenerate levels are split into $N$ spin doublets. The energy 
level of  each orbital state is specified as
\begin{eqnarray}
&& \varepsilon_{\sigma l}=\varepsilon_c+l\Delta_{orb} \\
&& \quad (l=-\frac{N-1}{2},-\frac{N-3}{2}, \cdots ,\frac{N-1}{2}) \nonumber
\end{eqnarray}
where $\varepsilon_c$ denotes the centre of the energy levels
and $l$($\sigma$) represents orbital(spin) index.

This type of orbital splitting has been experimentally realized 
as Fock-Darwin states in vertical quantum dot systems \cite{pap:Tarucha,pap:fds,pap:Tokura} or clockwise \& counter-clockwise states in carbon nanotube 
quantum dot systems \cite{pap:Herrero}, where the orbital 
splittings are proportional to the magnitude of magnetic fields.

\subsection{Bethe-ansatz exact solution }

We model the above quantum dot system by assuming that the Coulomb repulsion 
$U$ between electrons is sufficiently strong at the quantum dot, so that we 
effectively put $U\to \infty$.  We focus on the $N$-degenerate
 orbital states in the quantum dot, which are assumed to 
hybridize with the corresponding
 conduction
channels in the leads for simplicity.  In these assumptions, our system can be
described by an SU($N$) multi-orbital extension of the impurity Anderson model in the 
strong correlation limit ($U\to \infty$). The Hamiltonian reads
\begin{eqnarray}
H &=& H_0 + H_I, \\
H_0 &=& 
\sum_{\sigma, l}\varepsilon_{\sigma l} |\sigma l \rangle \langle \sigma l |
+ \sum_{k \sigma l} \varepsilon_k c_{k \sigma l}^{\dagger} c_{k \sigma l} , \\
H_I &=& \sum_{k \sigma l} (V_k |\sigma l \rangle \langle 0| c_{k \sigma l} + h.c.) ,
\label{hamil}
\end{eqnarray}
where $c_{k \sigma l}^{\dagger}(c_{k \sigma l}$) creates
(annihilates) a conduction electron with wavenumber $k$, 
spin $\sigma$$(=\pm1/2)$ and orbital $l$.  An electron 
state at the dot is 
expressed as $|\sigma l \rangle$, which is 
supplemented by the unoccupied state $|0 \rangle$.

In spite of its simple appearance, the Hamiltonian (\ref{hamil}) is 
difficult to treat generally 
because of strong correlations coming from  the $U\to \infty$
condition. Fortunately, this model was extensively studied 
two decades ago in the context of electron correlations 
for Ce and Yb rare-earth impurity systems. In particular,
 the exact solution of the model was obtained by
the Bethe ansatz method under the condition that the density of states 
for electrons in the leads is constant (wide-band limit)
 \cite{schkawa}. This solution enables us to exactly 
compute the static quantities, but not transport quantities in
a direct way. However, by
combining the Landauer formula and the Friedel sum rule, we can evaluate 
the zero-bias conductance 
 in the Kondo regime at absolute zero \cite{pap:Kawabata,pap:Meir},
\begin{eqnarray}
G=\frac{2e^2}{h} \sum_l \sin^2 \left( \pi n_{\sigma l} \right), \label{eq:conductanceLandaur}
\end{eqnarray}
where the occupation number of the electron $ n_{\sigma l}$ is given by
 the Bethe-ansatz solution.

\subsection{NCA method }

The above technique to calculate the conductance by the exact solution 
cannot be applied  to the finite-temperature case. Therefore, we 
complimentarily
 make use of the NCA 
  for the analysis at finite temperatures \cite{book:hewson,Bickers}.
The NCA is a self-consistent perturbation theory, which summarizes 
a specific series of expansions in the hybridization $V$. This method 
is known to give physically sensible results at temperatures around or
higher than the Kondo temperature. 
In fact, it was successfully applied to the Ce and Yb 
 impurity problem mentioned above, for which orbital 
degrees of freedom play an important role \cite{pap:Kuramoto,pap:Zhang,pap:Coleman,pap:Maekawa}.
Therefore, by combining  the exact 
results at zero temperature with the NCA results at
finite temperatures, we can 
deduce correct behavior of the conductance for the model (\ref{hamil}).

The NCA basic equations can be obtained in terms of coupled equations for two types of self-energies $\Sigma_0(z)$ and $ \Sigma_{\sigma l}(z)$ of the resolvents $R_{\alpha}(z) = 1/(z- \varepsilon_{\alpha} - \Sigma_{\alpha}(z))$,
\begin{eqnarray}
\Sigma_0(z) &=& \frac{2\Gamma}{\pi} \sum_l \int_D^D d\varepsilon \frac{f(\varepsilon)}{z-\varepsilon_{\sigma l}+ \varepsilon - \Sigma_{\sigma l}(z + \varepsilon)}, \\
\Sigma_{\sigma l}(z) &=& \frac{\Gamma}{\pi} \int_D^D d\varepsilon \frac{1-f(\varepsilon)}{z - \varepsilon_0+ \varepsilon - \Sigma_0(z + \varepsilon)},
\end{eqnarray}
for  the flat conduction band of width 2$D$.
At finite temperatures, the conductance is given as \cite{pap:Hettler},
\begin{eqnarray}
G = \frac{2e^2}{h} \Gamma \sum_l \int d\varepsilon \left( -\frac{df(\varepsilon)}{d\varepsilon} \right) A_{\sigma l}(\varepsilon),
\end{eqnarray}
where $f(\varepsilon)$ is the Fermi distribution function and 
$A_{\sigma l}(\varepsilon)$ is the one-particle spectral function including the effect of the above self-energies.

\section{Two-orbital Kondo effect}\label{sec:two-orbital}
We begin with a quantum dot system with two orbitals ($N=2$), where it is assumed  that the orbital degeneracy is lifted by a splitting $\Delta_{orb}$ while the spin degeneracy remains in the presence of magnetic field. 
As mentioned above, the two-orbital Kondo effect,
often referred to as the doublet-doublet or SU(4) Kondo effect,
has been experimentally observed in vertical dot systems and  carbon nanotube
dot systems \cite{pap:Sasaki2,pap:Herrero}.
We will be mainly concerned with the Kondo regime, where the characteristic
energy scale is given by the Kondo temperature $T_K^{SU(4)}$.

\subsection{Exact results at absolute zero}

\begin{figure}[bth]
\includegraphics[width=7cm]{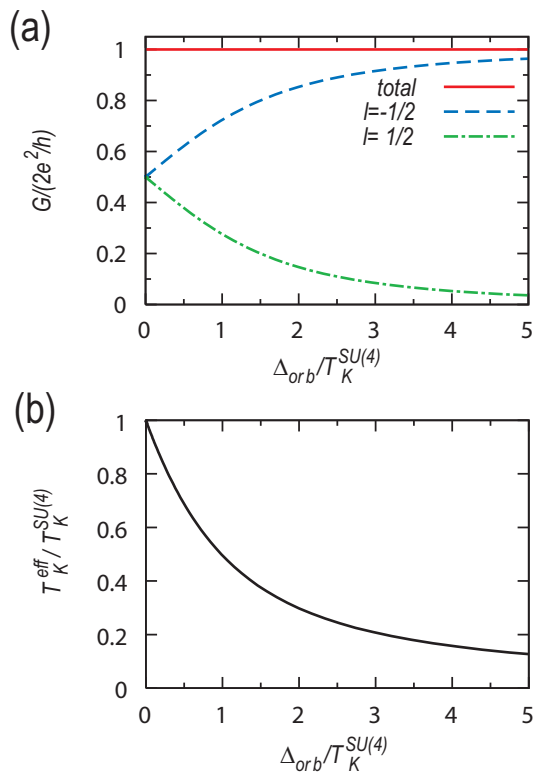}
\caption{(a) Zero-temperature conductance calculated by the exact solution as a function of the energy splitting for the SU(4) Anderson model in the Kondo regime. The total conductance is constant, although the contributions from two orbitals are changed monotonically. (b) The effective Kondo temperature $T_K^{eff}$. }
\label{fig:su4gkt}
\end{figure}
\begin{figure}[bth]
\includegraphics[width=7cm]{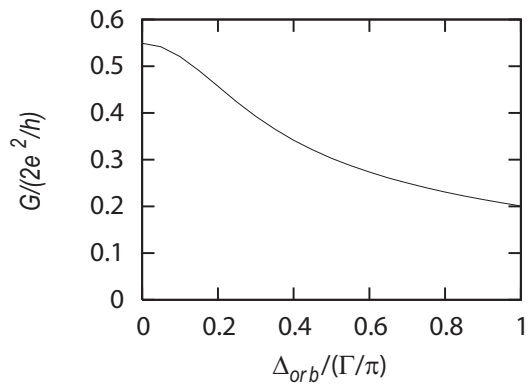}
\caption{Finite-temperature conductance computed by the NCA as a function of the energy splitting for the SU(4) model in the Kondo regime:
$k_B T = 0.1 \Gamma/\pi$ where $\Gamma$ is the bare resonance width.
The energy level of the dot is $\epsilon_c = -20 \Gamma / \pi$, which
gives the SU(4) Kondo temperature $T_K^{SU(4)}  \sim 0.1 \Gamma$.}
\label{fig:su4splt}
\end{figure}
It is known that  the conductance for the SU(4) case ($\Delta_{orb}=0$) and the SU(2) case ($\Delta_{orb}=\infty$) is both given by $2e^2/h$ at absolute zero. However, the conductance in the intermediate coupling region ($\Delta_{orb}= finite$) has not been discussed so far. The calculated conductance for each orbital is shown in Fig. \ref{fig:su4gkt}(a), which changes smoothly around the Kondo temperature $T_K^{SU(4)}$. An important point is that the total conductance is always $2e^2/h$, which is independent of magnetic fields. In fact, by putting the occupation number $n_{\sigma, \pm1/2} = 1/4 \mp \delta n(\Delta_{orb})$ into Eq. (\ref{eq:conductanceLandaur}), we can readily derive the constant conductance in the presence of magnetic fields. This result is characteristic of the SU(4) case, which is consistent with those deduced for the double-dot system with a special setup \cite{pap:Borda}.

It should be noted here that there is another important effect due to magnetic fields: the effective Kondo temperature changes with magnetic fields. The effective Kondo temperature, which is defined by the inverse of the spin susceptibility for the lowest doublet, gives the width of the effective Kondo resonance in the presence of the orbital splitting.
As  seen  in Fig. \ref{fig:su4gkt}(b), it monotonically decreases, and 
is inversely proportional to the orbital splitting in the asymptotic region: 
$T^{eff}_K \sim 1/\Delta_{orb}$ \cite{pap:Sakano,pap:EtoSU4,pap:sch3,pap:Yamada}.

\subsection{Analysis at finite temperatures}

It is remarkable that the conductance does not depend on magnetic fields at absolute zero, however, we should recall that the conductance is experimentally observed at finite temperatures. Since the Kondo temperature provides the characteristic energy scale of the system, the decrease of the effective Kondo temperature should have a tendency to suppress the Kondo effect at a given finite temperature. Therefore, we naturally expect from the zero-temperature analysis that the conductance monotonically decreases with increasing magnetic fields at a finite temperature. This can be confirmed  by the NCA numerical calculations at finite temperatures. The results are shown in Fig. \ref{fig:su4splt}, from which we indeed see the monotonic decrease of the conductance.
The obtained results  qualitatively agree with the experimental observation  of the SU(4) Kondo effect in the quantum dot systems \cite{pap:Sasaki2,pap:Herrero}. We wish to emphasize again that the magnetic-field dependence found here is different from that discussed so far for the ordinary spin Kondo effect without orbitals: the decrease of the conductance is caused by the decrease of the Kondo temperature in our scenario for the SU(4) case.

\section{Three-orbital Kondo effect}\label{sec:three-orbital}

In the previous section, we have considered the conductance for the two-orbital Kondo effect. We now generalize our treatment to the multiorbital cases with more than two orbitals. In contrast to the two-orbital case, where the conductance at finite temperatures shows behavior similar to the usual spin Kondo effect,
 we will see that the conductance for three- or more- orbital systems shows qualitatively different behavior in the magnetic-field dependence. This provides us with an efficient way to characterize the multi-orbital Kondo effect experimentally. Here, we discuss the characteristic transport properties  by taking the three-orbital Kondo effect ($N=3$) as an example.

\subsection{Exact results at absolute zero}
Exploiting the Bethe-ansatz exact solution and the Landauer formula, we calculate the conductance at $T=0$  for the single-dot system with three orbitals in the Kondo regime. We note that Schlottmann studied a similar model for Ce-impurity systems in a crystalline field to discuss  its magnetic properties \cite{pap:sch4}.
\begin{figure}[bth]
\includegraphics[width=7cm]{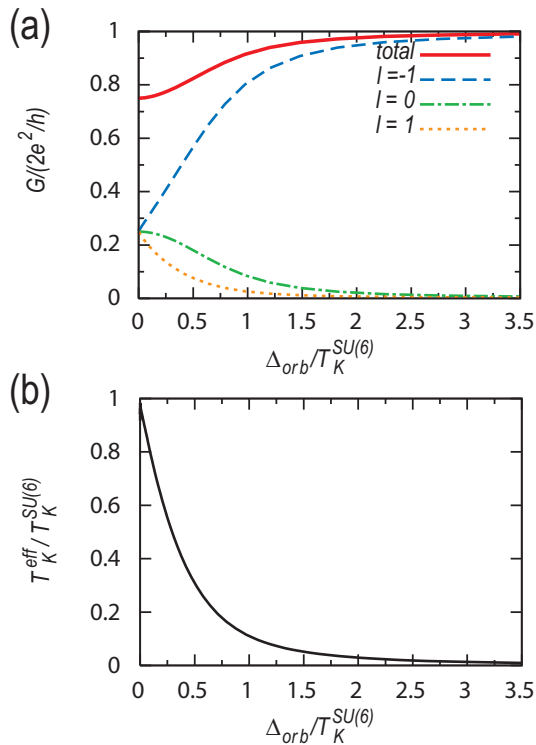}
\caption{(a) Conductance calculated by the exact solution at absolute zero as a function of the energy splitting $\Delta_{orb}$. Each line represents the total or the contributions from each orbital state. The system is in Kondo regime: the total number of the electrons in the quantum dot is $n_d=1$. (b) The effective Kondo temperature $T_K^{SU(6)}$.}
\label{fig:gn-d}
\end{figure}
The conductance computed by the exact solution is shown  as a function of the energy splitting  in Fig. \ref{fig:gn-d}.
It is seen that the total conductance at zero temperature monotonically increases from $3e^2/2h$ (SU(6) case) with the increase of energy splitting (or magnetic field), and approaches the value of $2e^2/h$ for large splittings (SU(2) case). It is noteworthy that the conductance is enhanced by the magnetic field in contrast to  the ordinary spin Kondo effect without orbital degrees of freedom, for which the conductance is 
simply suppressed by the magnetic field.
In more general cases with  $N$ orbitals, the conductance increases from $(2Ne^2/h)\sin^2 (\pi/(2N))$ to $2e^2/h$ with the increase of the magnetic field since the electron number of the lowest-orbital state  changes from  $1/(2N)$ (zero field) to $1/2$ (large fields). This analysis at zero temperature naturally leads us to predict the following behavior at finite temperatures. Since the magnetic field  decreases the effective Kondo temperature, there appears the competition between enhancement and suppression of the Kondo effect at finite temperatures. As a consequence, a maximum structure should appear in the conductance as a function of the magnetic field in general cases with more than two orbitals. This unique feature distinguishes the three- or more-orbital Kondo effect from  the two-orbital Kondo effect.

\subsection{Analysis at finite temperatures}

\begin{figure}[bth]
\includegraphics[width=7cm]{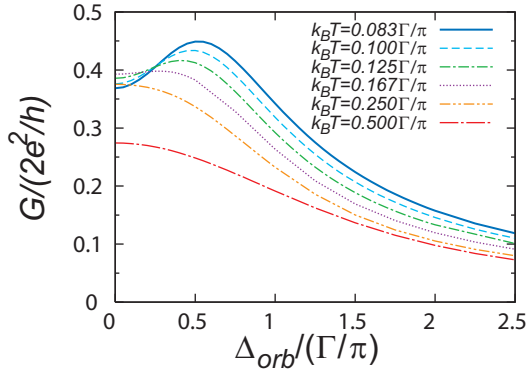}
\caption{Conductance for the three-orbital Kondo effect computed  by the NCA calculation at different temperatures as a function of the orbital splitting $\Delta_{orb}$. The parameter is $\varepsilon_c= -25\Gamma/\pi$, which gives $T_{K}^{SU(6)} \sim 0.2 \Gamma$. At low temperatures, the predicted maximum structure appears. Note that for $\Delta_{orb}=0$ and $k_BT=0.083\Gamma/\pi$, the total number of the electrons in the quantum dot is $n_d \sim 0.82$.}
\label{fig:ncaconductance}
\end{figure}

Here, we wish to confirm the conductance maximum at finite temperatures predicted from the zero-temperature analysis.  The conductance for the three-orbital Kondo effect computed  by the NCA at finite temperatures is shown in Fig. \ref{fig:ncaconductance}.
At low temperatures, the predicted maximum structure indeed emerges in the conductance as a function of the energy splitting. It is seen that the height of the maximum grows and its position gradually shifts toward the region with larger splitting $\Delta_{orb}$ as the temperature decreases. The shift reflects the fact that  the suppression of the Kondo effect due to the decrease of the Kondo temperature becomes somewat weaker at lower temperatures,.

\begin{figure}[bth]
\includegraphics[width=7cm]{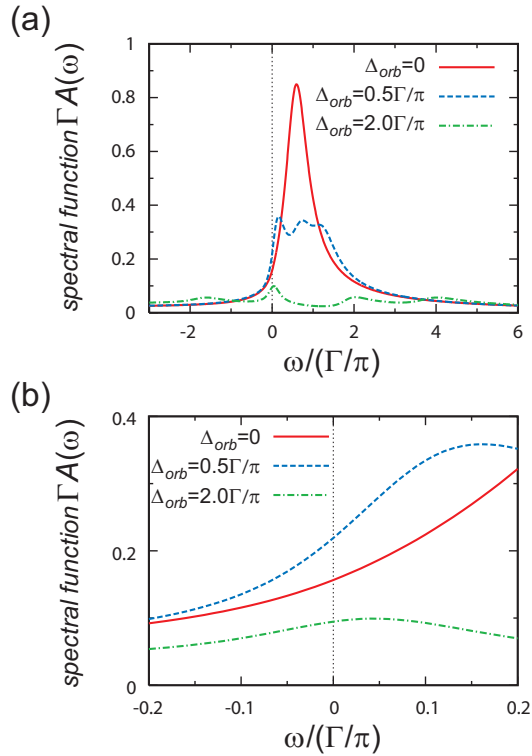}
\caption{(a)NCA results for the total spectral density $A(\omega)$ for different orbital splittings at $k_BT=0.083\Gamma/\pi$. (b)Zoom of $A(\omega)$ around the Fermi level of electrons in the leads $\omega=0$. The bare energy level is $\varepsilon_c=- 25\Gamma/\pi$.}
\label{fig:spectral}
\end{figure}

To see how the conductance maximum structure emerges for the three-orbital Kondo effect, we show the one-particle spectral density $A(\omega)$  computed at $k_BT=0.083\Gamma/\pi$ in Fig. \ref{fig:spectral}.
For the case of $\Delta_{orb}=0$ (three degenerate orbitals), there is a single SU(6) Kondo peak, while for $\Delta_{orb}=0.5\Gamma/\pi$, it splits into three peaks. For $\Delta_{orb}=2.0\Gamma/\pi$, we can see that the SU(2) Kondo peak is developed around the Fermi level $\omega=0$, for which only the lowest energy level ($l=-1$) is relevant. Note that further increase of the energy splitting shoud have a tendency to suppress the above SU(2) Kondo effect at a given temperature because of the decrease of the effective Kondo temperature.
The enlarged picture of the spectral density around the Fermi level is shown in Fig. \ref{fig:spectral}(b), which determines the linear conductance at low temperatures. It is seen that the spectral weight around the Fermi level once increases and then decreases with increasing energy splitting, which gives rise to the maximum structure in the conductance.

\begin{table*}[bth]
\caption{Comparison of the conductance among different Kondo effects 
 in the presence of magnetic field at zero and finite temperatures.}
\begin{ruledtabular}
\begin{tabular}{ccccc}
 temperature&spin Kondo effect & two-orbital Kondo effect& three- or more-orbital Kondo effect\\ \colrule
 $T=0$ &decrease&constant $2e^2/h$ & increase up to $2e^2/h$ \\
 $T \ne 0 $&decrease &decrease&maximum \\
\end{tabular}
\end{ruledtabular}
 \label{tbl:comp}
 \end{table*}

\section{Summary}\label{sec:conc}
We have discussed transport properties via the multi-orbital Kondo effect in a single quantum dot system  by using the Bethe-ansatz exact solution at zero temperature  and  the NCA at finite temperatures.
It has been shown that the orbital Kondo effect 
gives rise to some remarkable transport properties, 
which are different from the ordinary spin Kondo effect. 
The results are summarized in TABLE \ref{tbl:comp} in comparison with
 the usual SU(2) spin Kondo effect.

For the two-orbital SU(4) Kondo effect, it has been found that 
the conductance  at absolute zero
 is  constant ($2e^2/h$) irrespective of the strength
  of magnetic fields.  However, the effective Kondo temperature, which gives 
the energy scale at low-temperatures,  decreases monotonically. 
Therefore, if the conductance is observed at finite temperatures,
it decreases with the increase of the magnetic field,
in accordance with the recent experimental results for the conductance in the vertical quantum and the carbon nanotube quantum dot system \cite{pap:Sasaki2,pap:Herrero}. We stress again that although  the magnetic-field 
dependence at finite temperatures seems similar to the conventional
 SU(2) spin Kondo effect at first glance, the mechanism is different:
 the decrease of the conductance in the two-orbital SU(4) Kondo effect is
due to the decrease of the effective Kondo temperature. This fact should be 
properly taken into account for the detailed analysis of the field-dependent 
conductance due to the SU(4) Kondo effect.

In the three-orbital or more-orbital cases, the
conductance at absolute zero  increases in magnetic fields, 
and is then saturated  at the value of $2e^2/h$.  In this case also,
 the effective Kondo temperature decreases as the
field increases, so that the conductance features a maximum structure at 
finite temperatures.  Such unusual behavior in the conductance
 may be observed experimentally in vertical quantum dot  systems by
 tuning  multi-orbital degenerate states systematically \cite{amaha}.
 
In this paper, we have considered a simplified model for the quantum dot
with multiple orbitals: e.g. the tunneling matrix elements 
between the leads and the dot are assumed to be constant 
and independent of orbitals.  We believe that in spite of
such simplifications the present model captures essential 
properties inherent in the multi-orbital Kondo effect
at least qualitatively. It remains an interesting problem to
improve our model by incorporating the detailed structure of the quantum
dot parameters,  which is now under consideration.

\begin{acknowledgments}
We would like to express our sincere thanks to M. Eto and S. Amaha
for valuable discussions.
R.S. also thanks H. Yonehara for advices on computer programming,
and T. Ohashi and T. Kita for discussions.
\end{acknowledgments}


\end{document}